\documentclass[10pt,showpacs,nofootinbib,aps,superscriptaddress,eqsecnum,prd,notitlepage,showkeys]{revtex4-1}
\usepackage[latin9]{inputenc}
\usepackage{amsmath}
\usepackage{amssymb}
\usepackage{graphicx}
\usepackage{esint}
\usepackage{textcomp}
\usepackage{mathrsfs}
\usepackage{url}
\usepackage[unicode=true, bookmarks=false, breaklinks=false,pdfborder={0 0 1},backref=section,colorlinks=false]
 {hyperref}

\begin{document}

\title{Coda reconstruction from cross-correlation of a diffuse field on thin
elastic plates}

\author{Aida Hejazi Nooghabi}

\affiliation{Sorbonne Universit\'es, UPMC Univ Paris 06, CNRS, UMR 7193, Institut des Sciences de la Terre de Paris (ISTeP), F-75005 Paris, France}

\affiliation{ESPCI Paris, PSL Research University, Institut Langevin, 1 rue Jussieu, F-75005, Paris, France}

\author{Lapo Boschi}

\affiliation{Sorbonne Universit\'es, UPMC Univ Paris 06, CNRS, UMR 7193, Institut des Sciences de la Terre de Paris (ISTeP), F-75005 Paris, France}

\affiliation{Sorbonne Universit\'es, UPMC Univ Paris 06, CNRS, UMR 7190, Institut Jean Le Rond d'Alembert, \'equipe LAM, F-75005 Paris, France}

\author{Philippe Roux}

\affiliation{Laboratoire ISTERRE, Universit\'e Grenoble Alpes, CNRS, 1380 rue de la Piscine, F-38000, Grenoble, France}

\author{Julien de Rosny}

\affiliation{ESPCI Paris, PSL Research University, Institut Langevin, 1 rue Jussieu,
F-75005, Paris, France}
\begin{abstract}
This study contributes to the evaluation of the robustness and accuracy of
Green's function reconstruction from cross-correlation of strongly dispersed
reverberated signals, with disentangling of the respective roles of ballistic
and reverberated ('coda') contributions.
We conduct a suite of experiments on a highly reverberating thin duralumin
plate, where an approximately diffuse flexural wavefield is generated by taking advantage of the plate reverberation and wave dispersion. A large number of impulsive sources that cover the whole surface of the plate are used to validate ambient-noise theory through comparison of the causal and anticausal (i.e., positive- and negative-time) terms of the cross-correlation to one another, and to the directly measured Green's function. To quantify the
contribution of the ballistic and coda signals, the cross-correlation integral
is defined over different time windows of variable length, and the
accuracy of the reconstructed Green's function is studied
as a function of the initial and end times of the integral. We show that
even cross-correlations measured over limited time windows converge to 
a significant part of the Green's function. Convergence is achieved over a wide time window, which includes not only direct flexural-wave arrivals, but also the multiply reverberated coda. We propose a model, based
on normal-mode analysis that relates the similarity between the cross-correlation and the Green's function to the statistical properties of
the plate. We also determine quantitatively how incoherent noise degrades the estimation of the Green's function.
\end{abstract}
\maketitle

\section{Introduction}

In parallel with the striking demonstration that cross-correlation of diffuse ambient noise recorded at two points can lead to recovery
of the exact Green's function between the two receivers \cite{Weaver2001}, the spatio-temporal correlation between a set of receivers in the presence of an impulsive source has been investigated.
The results were particularly conclusive at the seismic scale,
where coda-wave interferometry \cite{Snieder2002} was used to monitor
highly scattering media with unprecedented accuracy. When averaged
over a significant number of regional earthquakes, the cross-correlation
of the coda waves also showed the possibility of recovering the Rayleigh
surface waves between seismic sensors spread across Mexico \cite{Campillo2003}.
The present paper aims to quantify the convergence of the correlation
function toward the Green's function at the laboratory scale in a
two-dimensional thin-plate configuration. This property is derived from the formulation of the Ward identity when the propagation medium is covered with incoherent sources. When generalized to continuous ambient noise (instead of
one or a set of impulsive sources), the cross-correlation theorem
has resulted in both imaging and monitoring applications in a variety
of fields, such as ultrasonics (e.g., \cite{Lobkis2001}), helioseismology
(e.g., \cite{Rickett1999}), ocean acoustics (e.g., \cite{Roux2004}),
seismology from the regional to the global scale (e.g., \cite{Campillo2014,Boschi2015}),
medical imaging (e.g., \cite{Sabra2007}), and structural engineering (e.g.,
\cite{Sabra2008}).

If the Green's function of the medium
is to be reconstructed using this method, the diffuse incident field
should result from the superposition of either decorrelated isotropic
plane waves \cite{Wapenaar2005} or uniformly distributed noise sources
\cite{Weaver2004}. While the first condition here is only valid in open
media, the second one assumes that the anelastic
attenuation is uniform within the medium of propagation.
In both cases, it can be
shown that the time derivative of the field cross-correlations between
two points coincides with the difference between the 'causal'
and 'anticausal' impulse responses; i.e., the Green's function between the two
points.

In practice, these conditions are only partially fulfilled. The mismatch
between the cross-correlation and the Green's function for a real noise source distribution
is a measure of the accuracy of the method. The mismatch in homogeneous
open media can be straightforwardly deduced from geometrical considerations
of the source distribution \cite{Roux2005}. In complex random media,
only statistical quantities can be inferred, and a mismatch appears
as a fluctuation of the cross-correlation. In the case of multiple scattering media, these fluctuations can be relatively well predicted from a shot-noise model \cite{Larose2008}.
However, only the use of multiple-scattering wave theory leads to fully
consistent results \cite{Tiggelen2003,Rosny2013}. 

Only a few studies have been devoted to the interferometric reconstruction of
Green's functions in bounded media. The related theory is, however, relevant to 
room acoustics \cite{Nowakowski2015} and passive
structural health monitoring \cite{Sabra2008,Chehami2015}. In thin plates,
the dispersion relation for noise generated by an air jet has
been recovered \cite{Larose2007}. More recently, an experimental
and numerical study was also conducted to determine the role of
the ballistic and coda part on the cross-correlation when the source distribution
was circular and uniform\cite{Colombi2014}.
Our study here follows Duroux et al. \cite{Duroux2010}, who investigated
the convergence of the cross-correlation of vibrational waves in a thin aluminum plate. A heuristic shot-noise model introduced by Larose et al. \cite{Larose2008}
was used to interpret the results. 

Here, similar to \cite{Duroux2010}, the use of a Doppler vibrometer
and the reciprocity theorem allows a very flexible study of the effects
of the source distribution on the cross-correlation. The set-up allows the comparison of the cross-correlation result with the transient response that is estimated with an active source. We also study
the symmetry of the correlation, because this is often an experimental
proxy for the convergence toward the Green's function. 

We also cross-correlate a part of the multiply
reflected wave ('coda') instead of the entire reverberating time. The
results are analyzed through modal decomposition of the cross-correlation over the eigen vibrational modes of the plate. This modal approach leads to
consistent modeling of the cross-correlation that is governed by different regimes. When the number of sources is large enough, the model shows that even with a very short part of the coda, a perfect estimation of the Green's
function can be obtained. On the other hand, the similarity
coefficient with a single source can be as small as a factor of $1/\sqrt{3}$ when the entire signal is correlated. The effects of uncorrelated noise (e.g.,
electronic noise) are also analyzed.

This paper is structured as follows: section II introduces
the cross-correlation for flexural waves, and section III describes the experimental set-up and
how the cross-correlation is built using the reciprocity theorem. In
section IV, a preliminary study is proposed in terms of the symmetry
of the Green's function reconstruction, and relates this to the signal-to-noise ratio.
In section V, the quantitative similarity between the direct impulse
response and that obtained via cross-correlation of passive signals is studied.
Section VI is devoted to an estimation of the similarity coefficient
when cross-correlation is performed on only a part of the coda and a finite number of sources. Finally, the effects of uncorrelated
noise on the correlation process are taken into account.

\section{Interferometry with Lamb waves}

The cross-correlation theorem holds under the hypothesis of a diffuse
wavefield. We assume here the definition of 'diffuse' as given by
\cite{Kinsler1999}: a wavefield is diffuse if all of the propagation directions
have equal probability. This is approximately achieved in practice
if the sources are distributed with equal probability with respect to
the position, and/or if the recordings are very long in time in the presence
of scattering and/or reverberation. \\
 In the present experiments, we record the guided plate waves, which are known as Lamb waves.
Lamb waves propagate according to two modes: symmetric and
antisymmetric \cite{Royer2000}. At low frequency,
the fundamental antisymmetric and dispersive Lamb mode ($A_{0}$) dominates the vertical plate displacements. According to the Kirchhoff-Love hypothesis\cite{Ventsel2001},
the Green's function ($G$) of the $A_{0}$ or flexural mode is the solution to
the equation of motion associated with an impulsive point source,
\begin{equation}
D{\Delta^{2}}G(\textbf{r}-\textbf{r}_{0},t)+{\rho_{s}}(\frac{\partial^{2}G(\textbf{r}-\textbf{r}_{0},t)}{\partial{t}^{2}}+\frac{1}{\tau_{a}}\frac{\partial{G(\textbf{r}-\textbf{r}_{0},t)}}{\partial{t}})=-\delta(\textbf{r}_{0})\delta(t),\label{eq:Geen}
\end{equation}
where $\Delta^{2}$ is the biLaplacian operator that is defined as the squared
Laplacian, ${\rho_{s}}$ is the surface density of the plate, $\tau_{a}$
is the attenuation time, \textbf{r} is the position vector, $t$ is the time, and $D=h^{3}E/12(1-\nu^{2})$ is
the bending stiffness, where $E$ and $\nu$ are the Young's modulus
and Poisson's ratio, respectively, and $h$ is the thickness of the
plate \cite{Fahy2007}. If the signal is generated by $N$ point sources
of identical power spectral density $S(\omega)$, the Fourier transform of the
cross-correlation estimated from a record of duration $\Delta T$ is given by
\begin{equation}
C(\textbf{r}_{l}^{R},\textbf{r}_{l'}^{R},\omega)=\Delta T \sum_{k=1}^{N}G(\textbf{r}_{l}^{R},\textbf{r}_{k}^{S},\omega)G^{*}(\textbf{r}_{l'}^{R},\textbf{r}_{k}^{S},\omega)S(\omega),
\end{equation}
where $G^{*}$ is the complex conjugate of $G$, $\textbf{r}_{l}^{R}$
and $\textbf{r}_{k}^{S}$ are the $l$-th receiver position and the $k$-th
source position, respectively. According to \cite{Chehami2014}, the cross-correlation is related to the imaginary part of the Green's function by 
\begin{equation}
C(\textbf{r}_{l}^{R},\textbf{r}_{l'}^{R},\omega)=\frac{\Delta T S(\omega){N\tau_{a}}}{\rho_{s}\omega A}ImG{(\textbf{r}_{l}^{R},\textbf{r}_{l'}^{R},\omega)}+Q(\omega)\label{correlG}
\end{equation}
where $A$ is the plate area, and the deviation $Q(\omega)$ can be caused, in particular, by non-uniformities
in the source distribution, as well as by instrumental error. Equation (\ref{correlG}) shows that the Green's function associated with two locations $\textbf{r}_{l}^{R}$ and $\textbf{r}_{l'}^{R}$ where sensors are deployed can be reconstructed by cross-correlation of the ambient recordings made by
the two sensors, provided that the ambient noise is diffuse \cite{Chehami2014}.

\section{Plate experiments}

\subsection{Set-up}

\begin{figure}[h!]
\includegraphics[width=8cm]{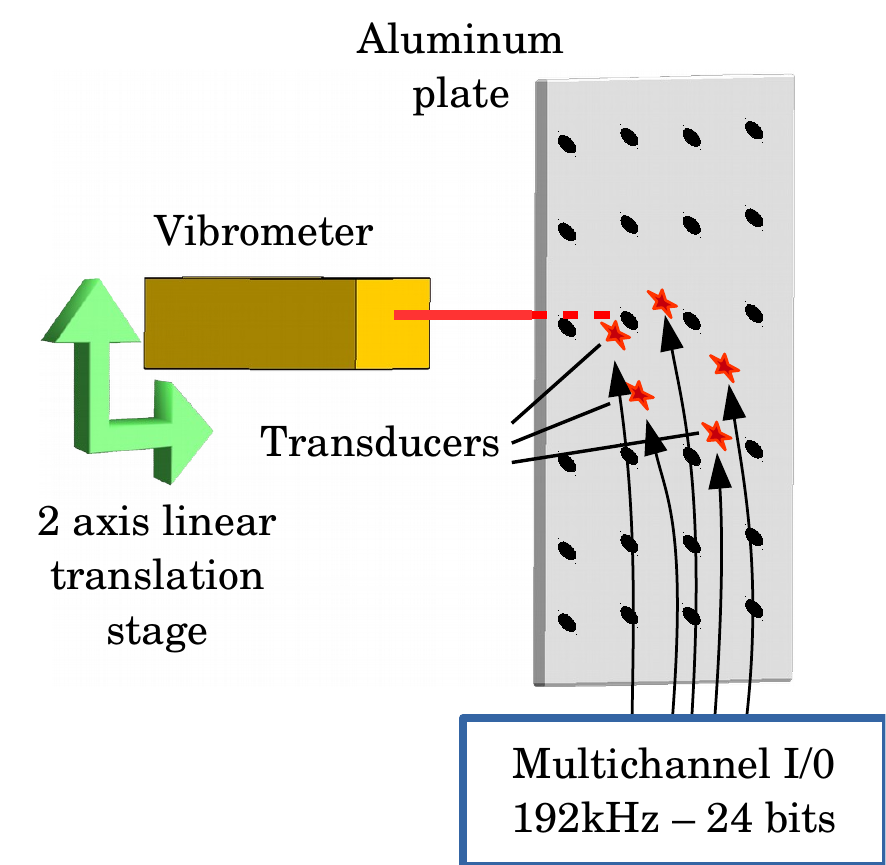} \centering \caption{Experimental set-up, consisting of a duralumin plate and five piezoelectric
transducers. The normal displacement of the plate is measured using
a laser vibrometer placed on a two-dimensional motor and at the points marked by the black circles. }
\label{Experimental_setup} 
\end{figure}

The experimental set-up consists of a homogeneous duralumin plate of 50 by 60 by
0.3 cm, with five piezoelectric transducers attached to the plate
using instant glue (Fig. \ref{Experimental_setup}). There are neither significant material 
heterogeneities within the plate, nor scattering obstacles attached to it. The temperature of the room is controlled by an air conditioning system. We measured the room temperature
continuously during the experiment and no significant variation was observed.
All of the transducers
are located at a minimum distance of 10 cm from the sides of the plate.
Instead of a short pulse, a linear 'chirp' in the frequency band of 100 Hz
to 40,000 Hz with the rate of frequency change of 39900 Hz/s is sent via these transducers. The transducers are made of a 200$\mu$m-thick ceramic piezoelectric disk adhered 
to a thin 20mm-diameter and 200$\mu$m-thick metallic disk. Effect of such a transducer on the propagation is limited because  total thickness of the transducer is only 13\% of the plate
thickness (the mass of the transducer equals 0.5g). On the other side of the plate, the vertical displacement induced by the vibration 
(i.e., Lamb waves) is scanned with a laser vibrometer.
Since the higher Lamb modes ($A_{1}$,$S_{1}$, $A_{2}$, $S_{2}$, ...) have a minimum cut-off frequency of 500 KHz, 
within our frequency band of interest, we can only generate the fundamental modes $A_{0}$, $S_{0}$ and $SH$ Lamb modes. Moreover, because we are working at less than a tenth of 
the cut-off frequency, the displacements recorded by the vibrometer are dominated by the $A_{0}$ mode. The latter, dubbed ``flexural-plate mode'' in the field of structural acoustics is very dispersive. Indeed,
between 100 Hz to 40000 Hz, the phase velocity ranges between 17 m/s to 1000 m/s (i.e., wavelength between 170 to 25 mm). The group velocity has almost twice the value of the phase velocity. At such low frequency 
range, propagation of the $A_{0}$ mode is well described by Eq.~ (\ref{eq:Geen}). Note that in 
this frequency range, Kirchhoff-Love model holds. 

Recorded signal is cross-correlated with the chirp
to extract the impulse response with the best signal-to-noise ratio.

\subsection{Fundamental properties of the observed signal}

\begin{figure}[h!]
\includegraphics[bb=1cm 0.cm 28cm 22cm,clip,width=10cm,height=14cm,keepaspectratio]{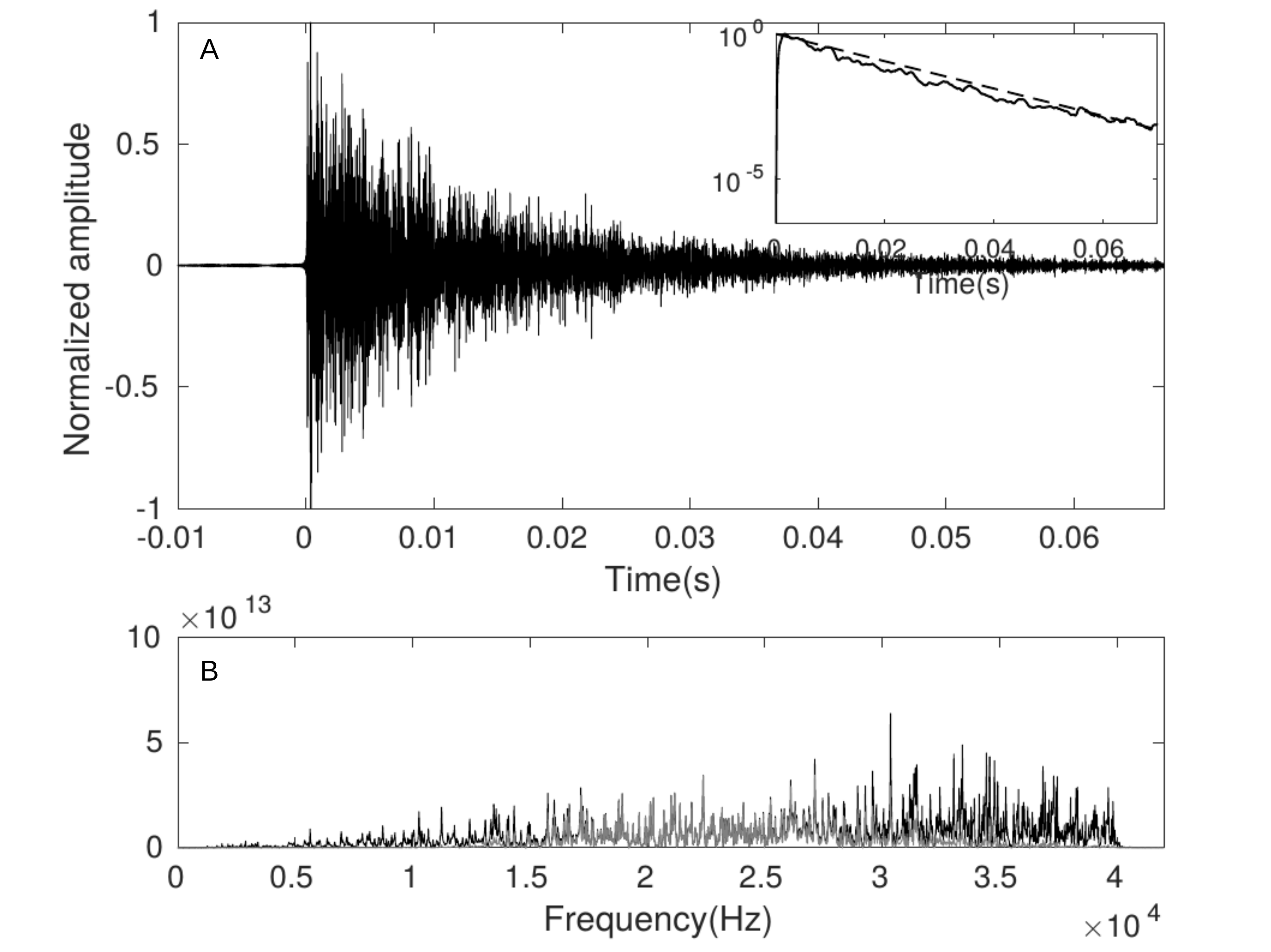}
\centering \caption{(A) Example of the Lamb waves recorded by a transducer after pulse compression of the emitted chirp signal. The source-receiver distance is about 21 cm. The inset illustrates the envelope of
the intensity of the signal (solid line) and the exponential fit according to $exp(-t /\tau_a)$ (dashed line), as a semilog scale. The attenuation time ($\tau_a$) is  9 ms. Data at negative time gives
a sense of the noise level. (B) Power spectrum of the raw signal (black) and bandpass filtered one (gray). The data is filtered between 15 KHz and 30 KHz using a Butterworth filter of 4th order. 
}
\label{sample} 
\end{figure}

Laser vibrometer scans are performed on a grid of about 2,500 regularly
spaced locations (with a uniform distance of 1 cm between two points
along the x and y plate directions), as qualitatively illustrated in Figure
\ref{Experimental_setup}. An example of the recorded signals after
chirp pulse compression is shown in Figure \ref{sample}. Based on the
reciprocity theorem (e.g., \cite{Aki1980}), from now on
we think of our readings as virtually emitted at the locations scanned
by the vibrometer, and virtually recorded at transducer locations.
This allows the simulation of the effects of a wide variety of source
distributions.

\begin{figure}[!h]
\includegraphics[bb=0.5cm 1cm 27cm 21cm,clip,width=14cm,height=12cm]{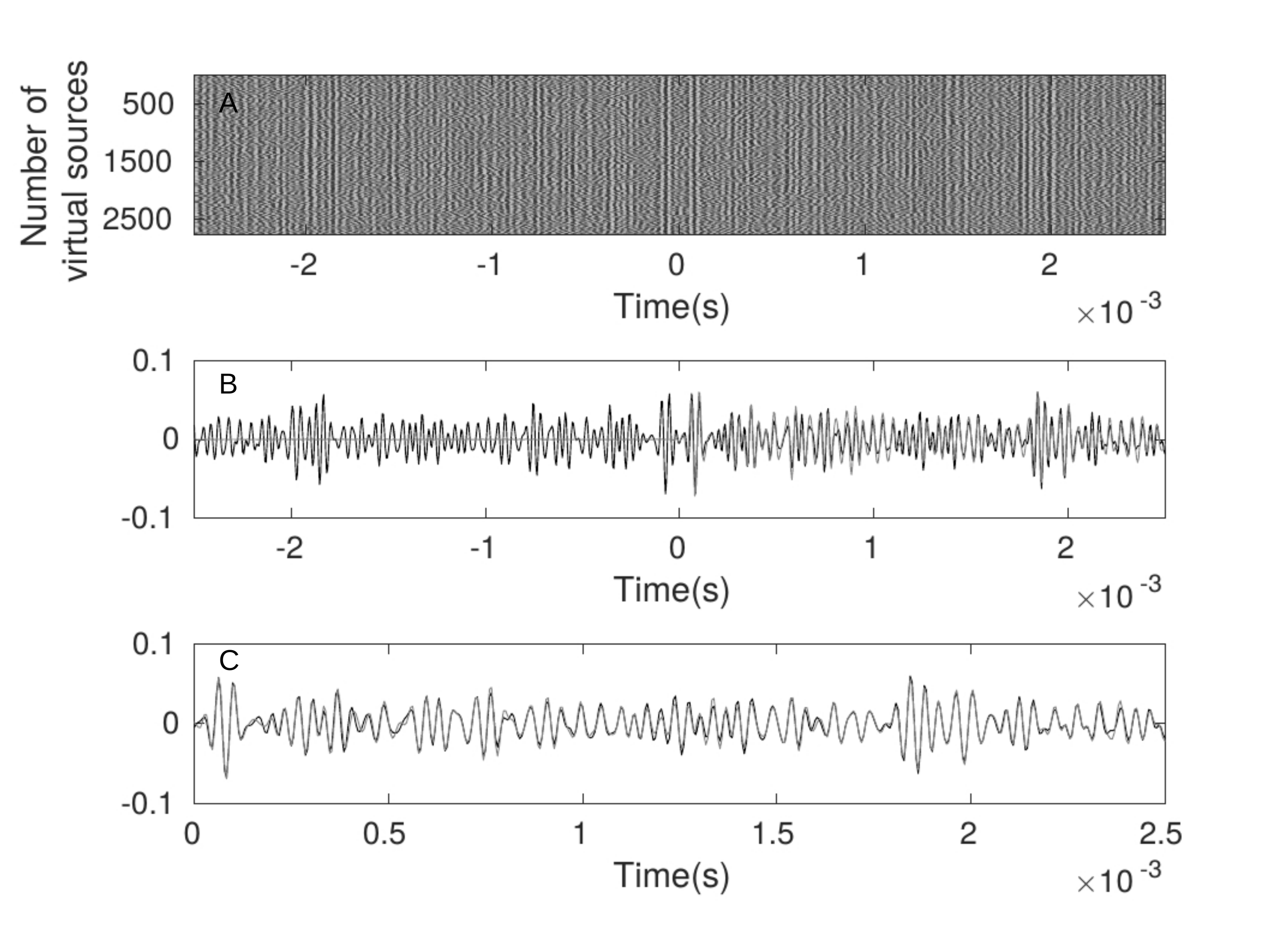}
\caption{(A) Cross-correlation gather for all of the available virtual point sources
(2,773 points) distributed uniformly on the plate. (B) Comparison of
the normalized cross-correlation averaged over all of the source positions
(black), and direct measurement of the source-to-receiver \char`\"{}time-integrated\char`\"{}
impulse response (gray). (C) Comparison between the positive time
(black) and the (flipped) negative time (gray) contributions
of the averaged cross-correlations. All of the cross-correlations in this plot are bandpass filtered by a Butterworth filter of 4th order between 15 KHz and 30 KHz, and normalization is performed
with respect to the maximum energy of the correlated signals.}
\label{uniformcc} 
\end{figure}

The envelope of the multi-reverberated impulse response shows exponential
decay. By fitting an exponential function to the squared signal (intensity), an attenuation time (time that it takes for the amplitude to decrease by 1/e of the maximum amplitude) of 9 ms is found. At 10 KHz, this time corresponds to about 18 reflections off the boundaries. For a single
pair of receivers separated by 12 cm and placed at minimum distance of about 19 cm from the borders of the plate, the cross-correlations for all of the virtual noise
source positions are gathered in Figure \ref{uniformcc}A, while Figure
\ref{uniformcc}B shows the results of the averaging over all of the source positions.
A coherent structure clearly appears in Figure \ref{uniformcc}A that
appears to be symmetric in time. This is confirmed in Figure \ref{uniformcc}B
and \ref{uniformcc}C, which are remarkably symmetric, including several
reverberated arrivals after the direct wave. 

As transducers are reciprocal devices, they can also be efficiently
used as emitters. What we measure is the convolution of the 
electro-elastic response of the transducer that acts as the source, the Green's function, the electro-elastic
response of the transducer that acts as the reciever, and the emitted signal.
Note that this $G(t)$ is the vertical-vertical component of the Green's function. We assume that within 
the frequency band of interest, the frequency responses of the transducers are flat, and hence
the impulse response is obtained by applying pulse compression of the emitted signals on the recordings.
 Hence, the impulse response can be measured directly
between the pair of transducers.
The time-integration result of this directly measured response after
pulse compression ($G(t)$) of the emitted signal is compared
with the cross-correlation, to validate the theoretical result (Eq. (\ref{correlG})), as discussed
in section II. There is excellent agreement between the two, even for
multiply reflected contributions, as shown in Figure \ref{uniformcc}B
(similar to the results of \cite{Mikesell2012}). This impulse response 
corresponds to the vertical component of the Green's function.
See Supplemental Material (part IV) at [URL] for the location of the transducers on the plate (\cite{supple}).

\section{Symmetry of the cross-correlation of the observed data}\label{sec:symmetry}

We infer from Figure \ref{uniformcc} that
to a good approximation, the cross-correlation is symmetric with respect to time when averaged
over all of the available sources in the two-dimensional plate. Its symmetry in time is
often used as a signature of its convergence toward
the Green's function (e.g., \cite{Roux2005}). In this section, we quantitatively
study the dependence of the time symmetry of the cross-correlation on the number of sources. \\
To this end, $C_N(t)$ (i.e., the cross-correlation averaged over $N$ sources) can be written as the sum of two terms: 
\begin{equation}
C_{N}(t)=C_{N}^{+}(t)+C_{N}^{-}(t),\label{symmetry-separation}
\end{equation}
where 
\begin{equation}
C_{N}^{+}(t)=\frac{C_{N}(t)+C_{N}(-t)}{2}\textrm{ and }C_{N}^{-}(t)=\frac{C_{N}(t)-C_{N}(-t)}{2},\label{oddeven}
\end{equation}
where for the sake of simplicity, $\textbf{r}_{l}^{R}$ and $\textbf{r}_{l'}^{R}$ have been dropped from the argument of $C_{N}$. 

By definition, $C_{N}^{+}(t)$ (resp. $C_{N}^{-}(t)$) is symmetric
(resp. antisymmetric) with respect to time.

\begin{figure}[h!]
\includegraphics[bb=0bp 6cm 612bp 21cm,clip,width=8.5cm]{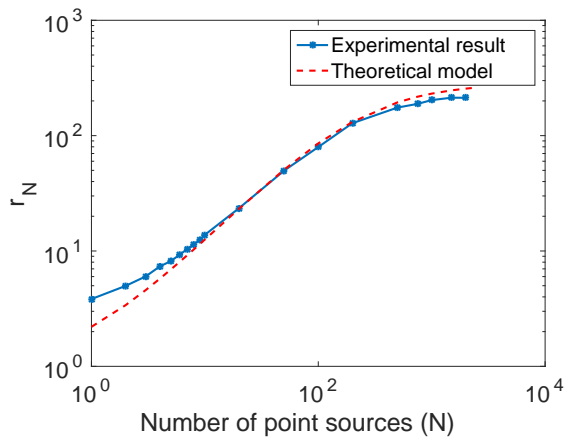}
\caption{\label{fig:Blue:--in} $r_{N}$ as a function of the number of sources $N$ picked randomly, derived from the experimental data (blue star line) using
Equation (\ref{eq:r}), and modeled (dashed red line) based on the theory developed in section~\ref{sec:symmetry} and summarized by Equation (\ref{eq:rf}). 
A 4th order Butterworth filter between 5000 Hz and 15000 Hz is applied to experimental data.}
\end{figure}

We then compute the ratio of the integrated squared $C_{N}^{+}(t)$
and $C_{N}^{-}(t)$; i.e., 
\begin{equation}
r_{N}\triangleq\frac{\int_{-\infty}^{+\infty}C_{N}^{+}(t)^{2}dt}{\int_{-\infty}^{+\infty}C_{N}^{-}(t)^{2}dt}.\label{eq:r}
\end{equation}
In Figure~\ref{fig:Blue:--in}, the experimental estimation of this
symmetry ratio is plotted with respect to the number of sources used
in the averaging process. This curve is determined by repeatedly implementing
Equation (\ref{eq:r}), based on a growing number of randomly distributed
sources, until all of the available sources are taken into account.

From Figure~\ref{fig:Blue:--in}, it appears that the cross-correlation is far from symmetric if only a limited number of sources are used in the averaging
process. In this case, the assumptions that allowed the derivation of Equation
(\ref{correlG}) do not hold. Nevertheless, as the number of sources
grows, the averaged cross-correlation becomes more and more symmetric in time. This
behavior is directly related to the convergence of the Green's function reconstruction. Let us first write the cross-correlation obtained for a subset $N$ of all of the point sources as 
\begin{equation}
C_{N}(t)=C_{\infty}(t)+\delta C_{N}(t),\label{CN}
\end{equation}
where $C_{\infty}(t)$ is the cross-correlation obtained when all of the recordings from all of the available sources in this experimental set-up are taken into account.

While $C_{\infty}(t)$ is the best possible approximation of the Green's function, given the experimental set-up,
there is no such thing as a perfectly uniform source distribution and infinite source density
in practical applications. It follows that $C_{\infty}(t)$ does not exactly coincide with $ImG$ (see Eq.~(\ref{correlG})), and should in principle be slightly asymmetric with respect to time. 
After defining the symmetric and antisymmetric components of $C_{\infty}(t)$,
similar to Equation (\ref{oddeven}), and substituting Equation~(\ref{CN})
into Equation (\ref{oddeven}) and Equation (\ref{eq:r}), we find that
\begin{equation}
r_{N}=\frac{4\int_{-\infty}^{+\infty}C_{\infty}^{2}(t)dt+2\int_{-\infty}^{+\infty}\delta C_{N}^{2}(t)dt}{4\int_{-\infty}^{+\infty}(C_{\infty}^{-}(t))^{2}dt+2\int_{-\infty}^{+\infty}\delta C_{N}^{2}(t)dt},\label{rC}
\end{equation}
where $C_{\infty}^{-}(t)$ is the antisymmetric component of $C_{\infty}(t)$.
Equation (\ref{rC}) is directly related to the ratio of the noise level (L),
which is a measure of the quality of the reconstructed Green's function from a finite
number $N$ of sources compared to the one obtained over an infinite
number of sources introduced in \cite{Moulin2015}. This can be expressed
in terms of $C_{\infty}(t)$ and $\delta C_{N}(t)$ 
\begin{equation}
L=\frac{\int_{-\infty}^{+\infty}\delta C_{N}^{2}(t)dt}{\int_{-\infty}^{+\infty}C_{\infty}^{2}(t)dt}.
\end{equation}
Thus, Equation (\ref{rC}) can now be rewritten as 
\begin{equation}
r_{N}=\frac{2+L}{2\zeta+L}\label{rrnl}
\end{equation}
where $\zeta=\int(C_{\infty}^{-}(t))^{2}dt/\int C_{\infty}^{2}(t)dt$
is the relative degree of asymmetry of the cross-correlation function
when $N\rightarrow\infty$. For a rectangular plate, L is given
by \cite{Moulin2015}
\begin{equation}
L\approx\frac{\pi n_{0}}{N\tau_{a}},
\end{equation}
where $\tau_{a}$ is the attenuation time and $n_{0}$ is the plate modal
density (modes per unit of frequency $\omega$), which in turn depends on the area of the plate surface $A$,
the surface density $\rho_{s}$, and the bending stiffness $D$, through
the relationship\cite{Fahy2007}
\begin{equation}
n_{0}=\frac{A}{4\pi}\sqrt{\frac{\rho_{s}}{D}}.
\end{equation}
Then, defining $k$ as the ratio of the attenuation time to the modal density,
\begin{equation}\label{k}
k=\frac{\tau_{a}}{n_{0}},
\end{equation}
Equation (\ref{rrnl}) finally becomes 
\begin{equation}
r_{N}=\frac{1+2Nk/\pi}{1+2\zeta Nk/\pi}\label{eq:rf}
\end{equation}
By finding adequate values of $k$ and $\zeta$, the best-fit
is obtained at 1.9 and 0.0035, respectively, which results
in the dashed red curve in Figure~\ref{fig:Blue:--in}. On the other hand, the 
experimental values for $\tau_{a}$
and $n_{0}$ are 0.009 s and 0.005 s/rad, respectively, which leads to $k=1.8$, according to Equation \ref{k}.
This experimental value for $k$ and that obtained from the fitting 
are consistent. The relative degree of asymmetry  $\zeta$ can be explained as a consequence of the fact that the laser vibrometer is only sensitive to vertical displacement (mainly $A_{0}$ mode). 
However, to some extent, the transducer also excites in-plane components (mainly $S_{0}$ and $SH$ mode). As a result, the in-plane components modes do not contribute to the reconstruction of the Green's function.

\section{Similarity between impulse response and cross-correlation}

\begin{figure}[ht]
\includegraphics[bb=1cm 14cm 20cm 22cm,clip,width=12cm,height=5cm]{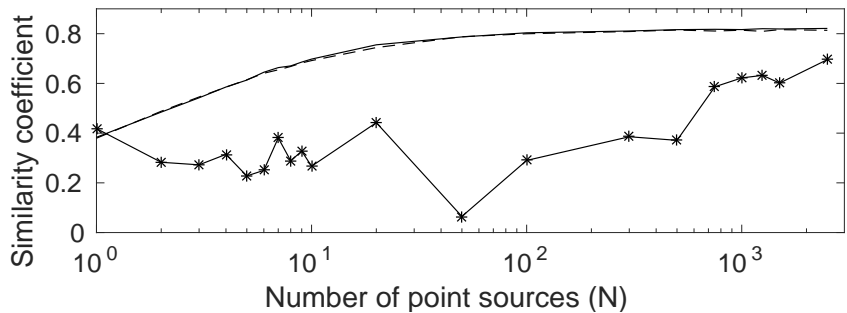}
\centering \caption{Similarity coefficient $S({G,C_{N}})$ between the $C_{N}(t)$ and $G(t)$ versus the number of point
sources (picked randomly) when the whole time signal is cross-correlated (solid line), and when only a part of the coda signal ($dT = 0.2$ s; see Eq. \ref{eq:cp}) is
cross-correlated (dashed line). A 4th order Butterworth filter between 15 KHz and 30 KHz is applied on $C_{N}(t)$ and $G(t)$. $\Delta\theta$ is equal to the attenuation time.
Curve with the star symbols is $S({G,C_{N}})$ where $C_{N}(t)$ is the result of the averaged cross-correlation of only direct-arrival window. In this case, $\Delta\theta$ is 0.7 ms.}
\label{compa} 
\end{figure}

The Pearson similarity coefficient 
between the two signals $G(t)$ and $C_{N}(t)$ is defined as (e.g., \cite{Press1992}),
\begin{equation}
S({G,C_{N}})=\frac{\left\langle \left|\int_{\Delta\theta}G(t)C_{N}(t)dt\right|\right\rangle }{\sqrt{\left\langle \int_{\Delta\theta}G(t)^{2}dt\right\rangle \left\langle \int_{\Delta\theta}C_{N}(t)^{2}dt\right\rangle }},\label{eq:similarity}
\end{equation}
where $\Delta\theta$ is the time-integration window to compute the
similarity coefficient, and $\langle \hdots \rangle$ is the averaging
over the different sets of point sources that are picked randomly. The latter is
required for $S({G,C_{N}})$ to be independent of the measurement positions. While
the word 'similarity' is preferred here
for clarity, $S$ is more often referred to in the literature as the 'correlation'
coefficient \cite{Press1992}. Figure~ \ref{compa} shows $S({G,C_{N}})$; i.e., the similarity coefficient between the
time-integrated impulse response and the cross-correlation of signals
recorded at transducers \#3 and \#4 for increasing number of sources.
The similarity coefficient $S({G,C_{N}})$ is computed over a time window with a length ($\Delta\theta$) 
equal to the decay time $\tau_{a}$ (see Fig. \ref{compa}).

To discriminate the contribution of the direct arrival, we have also plotted $S({G,C_{N}})$ where only direct arrivals are cross-correlated to get $C_{N}$. We clearly observe a lower similarity
unless there exists a very large number of sources. However, this contribution is negligible because when we cross-correlate the coda i.e., all the reverberated signal without direct arrival, the 
same value of the similarity coefficient is obtained as when all the signal is cross-correlated.
\\
First, for a large number of sources ($N>100$), $S({G,C_{N}})$ reaches a plateau
($s\approx0.82$) that is less than 1.0. 
From the analysis of the symmetry of the correlation, we measured that the latter is asymmetric with $\zeta=0.0035$. Assuming that the deviation to the perfectly reconstructed Green's function is as much symmetric 
as anti-symmetric, the total deviation reaches 0.0070. Consequently, the difference with the 18\% mismatch observed in Fig.~ \ref{compa} cannot be explained by the missing of in-plane components 
to the reconstructed Green's functions. We interpret this  increase of discrepancy by the fact that when considering the directly measured Green's function, the effect of the transducer response appears 
as the convolution of the two transducer responses, while when the recordings are cross-correlated, we deal with the 
cross-correlation of the transducer responses. A probably weaker effect comes from the absorption by the transducers that induces a deviation from the Green's function estimation as shown in \cite{Davy2016}.

Secondly, somewhat surprisingly, even with only one
or very few sources the cross-correlation matches 
the time-integrated Green's function relatively well ($S\approx0.4$), even when only a part of the coda is correlated.

In the next two sections, we propose models to explain these behaviors.
This begins with a study of the impact of the number of noise sources and
the length of the correlated coda on the emergence of the Green's function. Then, the effects of uncorrelated noise are taken into account.

\section{Correlation of windowed coda}

We next study the contribution of different parts of the coda signal
to the reconstruction of the Green's function. In practice,
rather than cross-correlating the entire signal, as above, the cross correlation is now limited to a time interval of variable length,
where the variable start-point and end-point are denoted as $T_{0}$ and $T_{0}+dT$, respectively. This is expressed by

\begin{equation}
C_{N}^{dT}(\textbf{r}_{l}^{R},\textbf{r}_{l'}^{R},t)=\sum_{k=1}^{N}\int_{T_{0}}^{T_{0}+dT}{G(\textbf{r}_{l}^{R},\textbf{r}_{k}^{S},\tau)G(\textbf{r}_{l'}^{R},\textbf{r}_{k}^{S},\tau-t)d\tau}.\label{eq:cp}
\end{equation}

Note that there is a distinction between the two time windows $dT$
and $\Delta\theta$ defined in Equation~(\ref{eq:similarity}); the former is the cross-correlation window, while
the latter is the time window over which the similarity
coefficient is computed.

\begin{figure}[h]
\includegraphics[bb=2cm 14cm 20cm 21cm,clip,width=12cm,height=5cm]{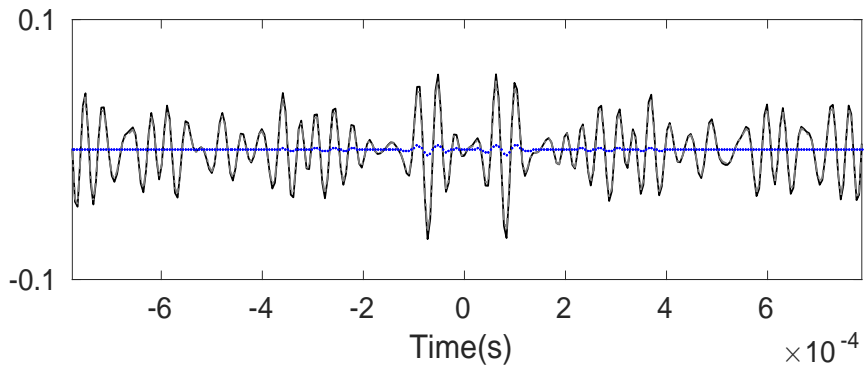}
\caption{Comparison of the cross-correlation results for three different time windows
when the sources are distributed over all of the plate, considering the whole
length of the signal (black line), only the ballistic part of the signal [$dT$ = 0.7 ms] (blue dots),
and only a part of the coda arrivals [$T_{0}$ = 0.4 ms, $dT$ = 0.02 s] (gray dashed line). All the cross-correlations
are bandpass filtered between 15 KHz and 30 KHz by a 4th order Butterworth filter.}
\label{Comparison-of-cross} 
\end{figure}

In Figure~\ref{Comparison-of-cross}, cross-correlations are compared when the time
window includes: only the ballistic wavefront, only the multi-reverberated
waves, and the whole wavefield. We observe that cross-correlation of only the ballistic part 
of the signals builds only the ballistic part of the Green's function, while 
cross-correlation of the coda part reconstructs both the ballistic and coda
parts of the Green's function. Moreover, the contribution
due to the ballistic path is negligible, and the recovery of the Green's function is dominated by the coda part.

To study the convergence process, the similarity coefficient is computed
between a reference cross-correlation $C_{ref}(\textbf{r}_{l}^{R},\textbf{r}_{l'}^{R},t)$
and the cross-correlation function obtained for different time
windows, averaged over randomly-picked sources $C_{N}^{dT}(\textbf{r}_{l}^{R},\textbf{r}_{l'}^{R},t)$. We
recall that $C_{ref}(\textbf{r}_{l}^{R},\textbf{r}_{l'}^{R},t)$
is obtained when all of the sources are emitting and
all of the signals are cross-correlated. In this section and the next one, the reference to obtain the similarity
coefficient is not the empirical Green's function ($G(t)$), but $C_{ref}(\textbf{r}_{l}^{R},\textbf{r}_{l'}^{R},t)$.
The reason for this is that according to Figure \ref{compa}, $G(t)$ never perfectly converges to $C_{ref}(\textbf{r}_{l}^{R},\textbf{r}_{l'}^{R},t)$, 
and they are not perfectly similar, which avoids reaching 1.0 for the similarity coefficient even 
when all of the sources are considered and the cross-correlation is over the whole signal.
Also, as explained before, this new choice for reference circumvents any influence on the results by
the difference in the frequency response of the transducers and the vibrometer.

\begin{figure}[h]
\includegraphics[bb=1cm 14cm 20cm 21cm,clip,width=30cm,height=4.5cm,keepaspectratio]{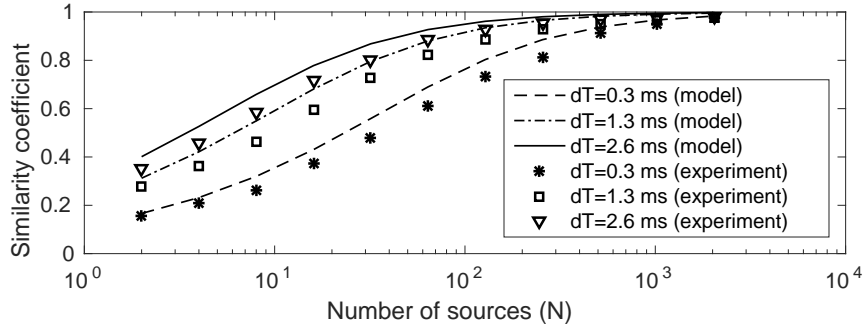}
\centering\caption{Similarity coefficient $S({C_{\infty},C_{N}^{dT}})$ between $C_{\infty}$ and $C_{N}^{dT}$ (cross-correlation obtained by
windowing one signal before the cross-correlation, and then averaged
over a subset of available sources picked at random). Symbols and lines show the experimental
results and theoretical model, respectively, for the various window lengths
($dT)$. The starting time ($T_{0}$) in all three cases is 2.3 ms,
i.e., no ballistic signal is cross-correlated.}
\label{fit_test3} 
\end{figure}

The similarity coefficient $S({C_{ref},C_{N}^{dT}})$ is computed in a window that contains the
direct arrivals and that lasts as long as the attenuation time, following Equation \ref{eq:similarity}. Figure \ref{fit_test3} shows the values of $S({C_{ref},C_{N}^{dT}})$ versus the number of point
sources ($N$) for the three different correlation window lengths ($dT$). It can be seen that all of the curves have a similar trend. 
The values of the similarity coefficient ($S({C_{ref},C_{N}^{dT}})$) increase with the number of sources ($N$),
and the resulting cross-correlation converges toward the reference one
after a certain number of sources, which indicates that the cross-correlation of
a short time window (compared to the decay time $\tau_a$ = 9 ms) leads to a reasonable impulse response.

See Supplemental Material (part I) at [URL] for our proposed model based on the modal decomposition
of the plate Green's function solution of Equation (\ref{eq:Geen}) to derive $S{({C_{ref},C_{N}^{dT}})}$ (\cite{supple}). 
Note that in the theoretical approach, we refer to the reference cross-correlation or the ``perfect'' 
cross-correction by ${C_{\infty}}$. This difference in the notation between the theoretical $({C_{\infty}})$ and experimental
$({C_{ref}})$ approach is a reminder for the fact that in the experimental case we cannot have an infinite number of sources 
and hence a perfect cross-correlation.
$S({C_{\infty},C_{N}^{dT}})$ is finally given by {\small{}
\begin{equation}
S({C_{\infty},C_{N}^{dT}})=\left[1+\frac{2}{N}\left(1+Z\right)\right]^{-0.5},\label{models}
\end{equation}
with

\begin{equation}
Z=\frac{\int\kappa(\delta\omega,\omega){\left|\widetilde{M}(\delta\omega)\right|^{2}}d\delta\omega}{\left|\widetilde{M}(0)\right|^{2} \left(1+F(\delta r)\right)}\label{eq:HH},
\end{equation}

where $F(\mathbf \delta r)$ accounts for the spatial correlation of the squared  eigen-modes,  $N$ is
the number of point sources, $\kappa(\delta\omega,\omega)$ is the two-level correlation and $\tilde M$ is the Fourier transform of the time windowed (the window duration being $dT$) squared mean intensity (see Supplemental Material for more details (\cite{supple})).

The two-level correlation function, that has been formally introduced in quantum chaos theory, is defined in terms of the modal density ($n_{0}(\omega)$) as
\begin{equation}
\kappa(\delta\omega,\omega)=\frac{\left\langle n_{0}(\omega)n_{0}(\omega+\delta\omega)\right\rangle }{\left\langle n_{0}(\omega)\right\rangle }.\label{eq:kappa_definiton}
\end{equation}
Lyon\cite{Lyon1969} also introduced it when analyzing the statistical properties of sound power in structures.

In a chaotic-shaped plate, due to repulsion between the modes, $\kappa$ is null when $\delta \omega = 0$ and is close to $n_0$ when $\delta \omega$ is larger than the modal density. The expression of $\kappa$ can be found in \cite{Mehta2004,Wright2010}.  
But here, as the plate geometry is regular, the eigen-frequency statistics follow a Poisson's distribution\cite{Lyon1969,Efetov1983}. As a consequence, there are no correlations between the eigenfrequencies, $\kappa(\delta\omega,\omega)$
is equal to the modal density and Z can be simplified into
\begin{equation}
Z=\frac{\pi n_0}{\tau_{a}\left(1+F(\delta r)\right)}\mathrm{coth}\left(\frac{dT}{\tau_{a}}\right)\label{eq:H}.
\end{equation}
The coefficient $Z$ can be interpreted as the average number of overlapping modes of the plate within the windowed cross-correlation at angular
frequency $\omega$.

See Supplemental Material (part III) at [URL] on derivation of the expressions for the spatial correlation of the squared eigenmodes and $F$ for chaotic and rectangular cavities (\cite{supple}). In the case of a chaotic cavity, based on the Berry conjecture\cite{Berry1977},
$F(\mathbf \delta r)$ is given by $2 J_{0}(k\delta r)^{2}$, where $J_{0}$ is the zero-th order Bessel function of the first kind. In the case of a rectangular cavity, using the same methodology as the one that led to the spatial correlation of the eigenmodes in a rectangular plate\cite{Bonilha1998}, we find 
\begin{equation}
 F(\mathbf \delta r) =\frac{J_{0}(2k\left|\delta x\right|)+J_{0}(2k\left|\delta y\right|)}{2}+\frac{J_{0}(2k\delta r)}{4}.
\end{equation}
Note that  now  the correlation of squared eigenmodes is anisotropic because it not only depends on modulus of $\delta \mathbf  r$, but also on the projection $dx$ (respect. dy) of $\delta \mathbf r$  along the x-axis (respect. y-axis).

The continuous curves in Figure
\ref{fit_test3} result from the model described by Equation (\ref{models}),
where $F(\delta r)\sim1/2$ for $\delta r\gg\lambda$ and $1+F(\delta r)$
is set to 1.5. We believe that this long-range correlation is due
to strong periodic orbits that are not taken into account by our correlation models. 

The convergence toward the Green's function is driven by three characteristic times: the modal density ($n_{0}$), the attenuation time of the plate ($\tau_{a}$); and the time window selected for the cross-correlation ($dT).$ Schematically,
three asymptotic cases can be identified: 
\begin{enumerate}
\item When $\tau_{a}\gg n_{0}$ and $dT\gg n_{0}$, the modes of the plate
are resolved because the attenuation is low (see Fig.~ \ref{fig:four-config}a).
Moreover, the integration time $dT$ is sufficiently large to include
a sufficient amount of coda to not induce modal overlapping by a windowing
effect. In such a case, the similarity coefficient $S({C_{\infty},C_{N}^{dT}})$ is high, and therefore the 'best' correlation is obtained. In the case of a single source $S({C_{\infty},C_{N}^{dT}})$ is $1/\sqrt3$.
\item When $\tau_{a}\lesssim n_{0}$ and $dT\gg\tau_{a}$, $Z\approx\pi n_{0}/\tau_{a}$,
the overlapping due to the attenuation is not negligible anymore (see
Fig.~ \ref{fig:four-config}b). The convergence of the cross-correlation
toward the Green's function is slower. Hence, at least Z sources are required to
obtain a good estimation of the Green's function from the cross-correlation.
\item Finally when, $dT\ll\tau_{a}$ and $dT\lesssim n_{0}$, the mode overlapping
is dominated by the effects of the coda truncation, and is given by
$Z\approx\pi n_{0}/dT$ (see Fig.~ \ref{fig:four-config}c). Again,
to get a good estimation of the Green's function, at least Z sources have to be used. 
\end{enumerate}
Hence, when $N$ is large compared to $1+Z$, $S({C_{\infty},C_{N}^{dT}})$
converges toward 1.0, independent of the window length $dT$. In other
words, even with a very short integration time in Equation~(\ref{eq:cp}),
the Green's function can be completely recovered. 

\begin{figure}[ht]
\includegraphics[bb=1cm 0cm 17cm 25cm,clip,width=50cm,height=10cm,keepaspectratio]{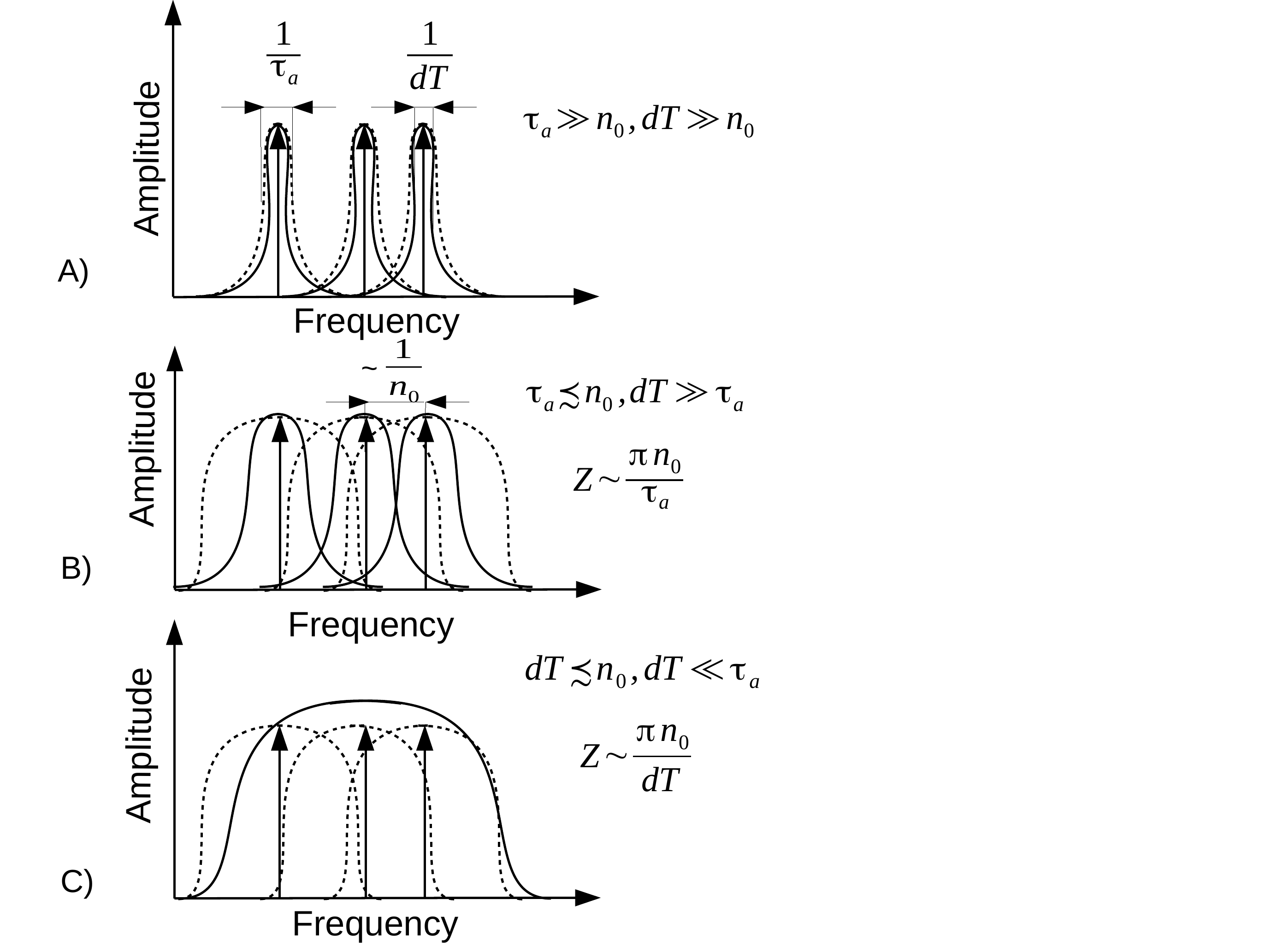}

\caption{\label{fig:four-config}Three schematic representations of a small
part of the Fourier transform of the window cross-correlations for
different values of the characteristic times of the system, namely
as: modal density ($n_{0}$), attenuation time ($\tau_{a}$), and selected time window for cross-correlation ($dT$). The vertical arrows
mark the eigen-modes that are separated on average by $\frac{1}{n_{0}}$.
The dashed curves represent Lorentzian spectra of the width $1/\tau_{a}$,
which is the inverse Fourier transform of the exponential attenuation.
Finally, the dotted curves show the sinc functions of the width 1/dT,
which is the Fourier transform of the window. The regimes and the corresponding
approximate values of $Z$ are denoted at the top righthand corner of each
plot.}
\end{figure}

Figure \ref{fit_test3} shows that the experimental results approximately
confirm this model; the similarity coefficient $S({C_{\infty},C_{N}^{dT}})$ does not converge to
1.0 exactly, which we ascribe to the presence of a small amount of random
noise in the measurements.

\section{Effects of random noise}

A comparison between the model and the experimental results (Fig.
\ref{fit_test3}) suggests that uncorrelated noise (e.g., electronic
noise) degrades the Green's function. We next analyze this effect by performing
the cross-correlation over time windows that start between $T_{0}=0$
and $T_{0}=57$ ms after the beginning of the transient response, and
end at time $T_{0}+dT=250$ ms (see inset in Fig.~\ref{test2_fartime}).
Time $T_{0}+dT$ is chosen to be a lot larger than $\tau_{a}$, to increase
the relative contribution of noise in the correlation process. The
similarity coefficient $S({C_{ref},C_{ref}^{dT}})$ is then retrieved according 
to Equation \ref{eq:similarity} and from the direct path and
primary coda signal ($\Delta\theta=2$ ms). As for the previous section,
the reference of the similarity coefficient is the one obtained when
the number of sources is large and the effects of noise are negligible.
Note that, in this case, all of the sources $N$ are considered in the computation
of $S({C_{ref},C_{ref}^{dT}})$, and the similarity coefficient is plotted 
in Figure~\ref{test2_fartime} as a function of $T_0$.

\begin{figure}[ht]
\includegraphics[bb=0.0cm 6.3cm 28cm 17cm,clip,width=30cm,height=5.5cm,keepaspectratio]{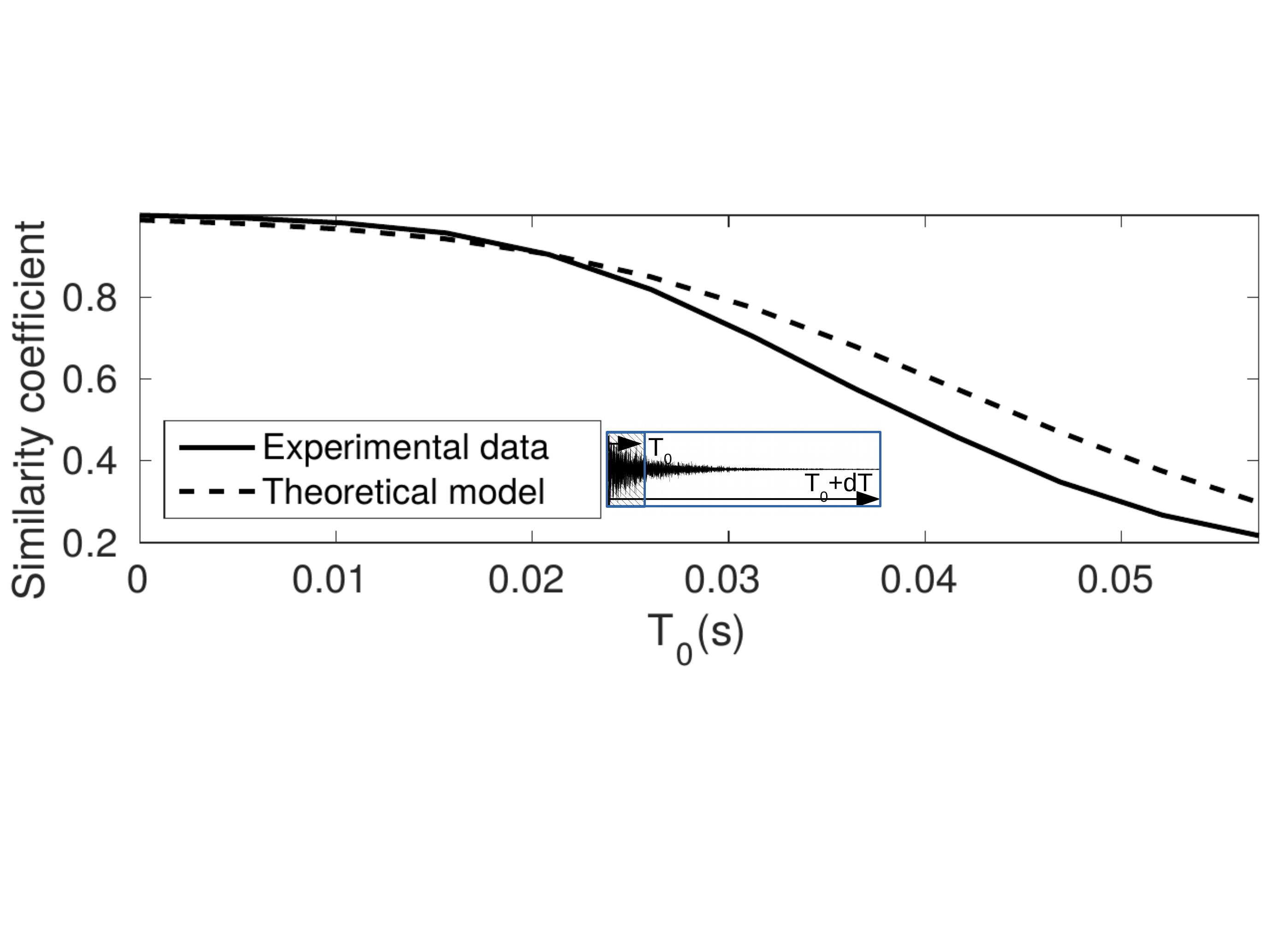}
\caption{$S({C_{\infty},C_{\infty}^{dT}})$ between $C_{\infty}(\textbf{r}_{l}^{R},\textbf{r}_{l'}^{R},t)$
and $C_{\infty}^{dT}(\textbf{r}_{l}^{R},\textbf{r}_{l'}^{R},t)$ given
by Equation~(\ref{eq:cp}). The end of the time windows ($T_{0}+dT)$
is fixed at 250 ms. $S({C_{\infty},C_{\infty}^{dT}})$ is then evaluated in a window that includes direct path and primary coda ($\Delta\theta=2$ ms).}
\label{test2_fartime} 
\end{figure}

We observe that the effects of noise becomes significant only when
$T_{0}$ is $>$15 ms. The decay of the $S({C_{ref},C_{ref}^{dT}})$
can be explained by assuming that the field measurement is perturbed
by a constant amount of uncorrelated noise at each transducer, according
to Equation (\ref{model_noise}). See Supplemental Material (part II) at [URL] 
for the complete derivation of $S({C_{\infty},C_{\infty}^{dT}})$ (\cite{supple}). Similar to the previous section,
we discriminate the experimental case from the theoretical one by changing ${C_{ref}}$ to ${C_{\infty}}$.
After some algebra, we find
\begin{equation}
S({C_{\infty},C_{\infty}^{dT}})\approx\left(1+\frac{\beta B(dT)}{N[e^{-T_{0}/\tau_{a}}(1-e^{-dT/\tau_{a}})]}\right)^{-0.5},\label{model_noise}
\end{equation}

where $B$ is the bandwidth of the system, and the dimensionless value $\beta$ is an indication of the noise-to-signal ratio.

The experimental data were fitted to Equation~(\ref{model_noise}).
We observe that the model explains the decay in $S({C_{ref},C_{ref}^{dT}})$
and as we go further in time, it is the noise content that dominates, 
which consequently causes degradation of the reconstructed
Green's function. Note that unlike the experimental curve, the theoretical one does
not start from the value of 1.0. This is due to a different reference in the 
similarity coefficient in these two cases. On the one
hand, for $T_{0}=0$, the correlation and the reference correlation
are identical because we chose 
the best correlation possible for the reference cross-correlation (i.e., at $T_{0}=0$). Consequently,
$S({C_{ref},C_{ref}^{dT}})$ is exactly 1.0. On the other hand, from a theoretical
point of view $S({C_{\infty},C_{\infty}^{dT}})$ is smaller than 1.0 because the reference
correlation is the ideal one, i.e., the difference between the advanced
and delayed Green's function, while the noise slightly contaminates the cross-correlation. 

\section{Discussion}

As explained by \cite{Derodeandlarose}, the process of reconstructing 
a Green's function by diffuse-field cross-correlation is closely related to 
that of acoustic time reversal \cite{Derode2003}. Accordingly, this study
has some implications in the context of time reversal.
To illustrate this point, let us assume that a pulse is sent at position
$\textbf{r}_{A}$ and the $N$ transient responses are recorded by
$N$ receivers at positions $\textbf{r}_{i}$. Each response is then
flipped in time and sent back into the medium by the $N$ emitters
located at the same positions as the $N$ receivers. In a reciprocal
medium, it can be formally shown that the time-dependent amplitude of
the time-reversed field at position $\textbf{r}_{B}$ is proportional
to the correlation $C_{N}(t)$ between positions $\textbf{r}_{A}$
and $\textbf{r}_{B}$. When only a part of the transient responses
between times $T_{0}$ and $T_{0}+dT$ are flipped and sent back,
the time-reversed field is then given by $C_{N}^{dT}(t)$. This formal
equivalence between the cross-correlation and the time reversal provides new insights \cite{Derode2003}.
First, when there is only one source, the cross-correlation behaves
approximately as the one-channel time-reversal signal of flexural
waves obtained by Draeger and Fink \cite{Draeger1997}. In particular,
it has been shown that in a lossless cavity, the time reversed field
at position $\mathbf{r}_{1}$ is equal to $G(\mathbf{r}_{1},\mathbf{r}_{1},t)\otimes G(\mathbf{r}_{A},\mathbf{r}_{B},-t)$ \cite{Draeger1999}.
For this set-up, we infer from Equations~(\ref{models}) and (\ref{eq:H})
that when $dT\gg\tau_{a}\gg n_{0}$, the similarity coefficient $S(C_{\infty},C_{1}^{dT})$
for one single source is equal to $1/\sqrt{3}$. Hence, the mismatch
is due to $G(\mathbf{r}_{1},\mathbf{r}_{1},t)$, i.e., fluctuations
induced by alternation of nodes and antinodes at $\mathbf{r}_{1}$.
Secondly, the case where the sources are uniformly distributed over
the surface corresponds to the instantaneous time reversal, which is also called
the Loschmidt echo \cite{Loschmidt1876}. In this case, a field can be perfectly time reversed at time $t_{0}$ by imposing
the initial condition $\psi(\mathbf{r},t_{0})$ and its negative time
derivative $-\partial\psi(\mathbf{r},t_{0})/\partial t$. Consequently,
by time reversing a very small window around time $t_{0}$ of the
transient responses by many sources over all of the surface, the time-reversed field is perfectly recovered for $t<t_{0}$. This result
is in agreement with our finding that $S(C_{\infty},C_{N>>1}^{dT})$ converges to 1.0 as $N$ grows.

A popular technique used in order to improve the estimation of the Green's function consists of correcting the exponential attenuation decay to artifically increase the attenuation time. In such a case, because $\tau_a$ is larger (see Eq.~(\ref{eq:rf})), the symmetry ratio  $r_N$ increases faster with the number of sources. The similarity coefficient is also improved because $\tau_a$ can be larger than the modal density (also called Heisenberg time) and the system behaves as if the plate modes are resolved (see Eq.~(\ref{models})). However, this correction also increases exponentially noise-to-signal ($\beta$) and therefore limits the efficiency of the method (see Eq.~(\ref{model_noise})).

In the case of continuous incoherent noise sources, the cross-correlation
converges toward the windowed cross-correlation, but with a window that
includes all of the transient signal. The present study provides new insights
into the understanding of the relative contributions of the different parts of the
transient response (e.g., ballistic and early coda, late coda) for
building the cross-correlation. Considering the transient recorded
signals that consist of direct and later arrivals, and cross-correlating
these parts of the signals separately, we show that exclusion of the 
direct arrivals and cross-correlation of only the coda
arrivals is very close to the case where we cross-correlate the full-time
signals.

\section{Conclusions}

Given any receiver pair, the signal that would be obtained at
either one from a source at the other can be reconstructed experimentally
by cross-correlation of the recordings of a diffuse field. The field can
be diffuse as a result of a dense, homogeneous source distribution
throughout the medium, and/or of scattering or reverberation:
this study was aimed at the disentangling of these two effects. We conducted
experiments with a thin plate where the surface was densely scanned by a
laser vibrometer, and where an array of transducers was deployed. This
set-up provides almost perfect control of the spatial distribution
of the transient sources. We first validated the theory through comparison of
the averaged cross-correlations and the directly observed Green's function. We also experimentally studied
the symmetry (with respect to time) of the cross-correlations,
as well as their similarity to the Green's function, as a function of the number
of uniformly distributed point sources. To explain
these observations quantitatively, a theoretical analytical model was developed that predicts the observed asymmetry of the averaged cross-correlation.
We next studied the convergence of the averaged cross-correlation
for time windows of variable lengths, which might be very short, and taken
at different points in the coda of the recordings. Here, a 
relatively dense/ uniform source distribution can result
in good estimation of the Green's function. We demonstrated that this
time window does not have to include the direct-arrival part of the signals
for the estimated Green's function to be a good approximation of the exact one.
Through statistical modal analysis, the respective
contributions of attenuation time, modal density, and number of sources
to the convergence of the cross-correlation toward the Green's
function were identified. Finally, we demonstrated both theoretically and experimentally that this convergence strongly depends on the position of the correlation time window only when additive random noise is taken into account. The relative effects of the noise are stronger when the late coda is cross-correlated. 

\section*{Acknowledgement}

This project received funding from the European Union Horizon
2020 research and innovation programme under Marie Sklodowska-Curie
grant agreement No 641943, and partial support from LABEX WIFI (Laboratory
of Excellence within the French Program \char`\"{}Investments for
the Future\char`\"{}), under references ANR-10-LABX-24 and ANR-10-IDEX-0001-02
PSL and ISterre within Labex@2020.

\bibliographystyle{unsrt}
\bibliography{biblio3}

 \end{document}